\documentclass[aps,prd,12pt,nofootinbib,tightenlines,bibnotes,superscriptaddress,notitlepage]{revtex4-1}
\usepackage{amsfonts,amssymb,amsmath,epsfig,bm}
\usepackage{graphicx}
\usepackage[usenames,dvipsnames]{xcolor}
\RequirePackage{color}
\RequirePackage[colorlinks=true,
urlcolor=Maroon,
anchorcolor=Maroon,
citecolor=Maroon,
filecolor=Maroon,
linkcolor=blue,
menucolor=Maroon,
linktocpage=true,
pdfproducer=medialab,
pdfa=true,
pdfusetitle
]{hyperref}

\usepackage{times}

\newcommand{\be}{\begin{equation}}
\newcommand{\ee}{\end{equation}}
\newcommand{\ba}{\begin{eqnarray}}
\newcommand{\ea}{\end{eqnarray}}
\newcommand{\bd}{\begin{displaymath}}
\newcommand{\ed}{\end{displaymath}}
\def\thalf{{\textstyle{\frac{1}{2}}}}

\begin{document}

\title{Randall--Sundrum Model with a Dilaton Field at Finite Temperature}
\author{Aditya Dhumuntarao}
\affiliation{School of Physics \& Astronomy, University of Minnesota, Minneapolis, MN 55455,USA}
\author{Joseph I. Kapusta}
\affiliation{School of Physics \& Astronomy, University of Minnesota, Minneapolis, MN 55455,USA}
\author{Christopher Plumberg}
\affiliation{Theoretical Particle Physics, Department of Astronomy and Theoretical Physics,
Lund University, S\"olvegatan 14A, SE-223 62 Lund, Sweden}
\vspace{.3cm}
\date{\today}

\parindent=20pt

\begin{abstract}
We find exact finite temperature solutions to $d=5$ Einstein-dilaton gravity with black branes and a Randall--Sundrum 3-brane. We show that there exists a unique generating superpotential for these models.   The location of the black brane and the associated Hawking temperature depend on the value of the the 3-brane tension while other parameters are held fixed. The thermodynamics of these solutions are presented, from which we show that the entropy satisfies $S<\text{Vol}(\mathbb{R}^3)/4G_5$. We demonstrate that in a certain limit the gauge dual of this theory effectively reduces to $\mathcal{N}=4$ SYM at finite temperature on $S^1\times\mathbb{R}^3$.
\end{abstract}

\maketitle

\section{Introduction}

The gauge/gravity conjecture has provided insights into the phenomenology of large $N_c$, strongly coupled gauge theories at finite temperature \cite{Maldacena:1997re,Gubser:1998bc,Witten:1998zw}. The holographic duality encodes finite temperature properties of $d$-dimensional gauge theories on $S^1\times\mathbb{R}^d$ (or $S^d$) into the thermodynamics of $(d+1)$-dimensional gravitational theories. In these non-perturbative regimes, the behavior of the strongly coupled plasma state may be inferred by calculating geometric properties of black brane (hole) solutions in the gravitational dual. 

The identification of a gravitational dual to Quantum Chromodynamics (QCD), the standard model, or extensions of the standard model would be a novel application of the conjecture and remains an open problem. A bottom-up approach to realize the holographic construction, known as AdS/QCD \cite{Karch:2002sh, Karch:2006pv, Erlich:2005qh}, deforms the boundary Conformal Field Theory (CFT) towards QCD by introducing 3-branes in conjunction with additional bulk fields. These ``soft wall" AdS/QCD models are used to reproduce certain phenomenological features such as electroweak or chiral symmetry breaking \cite{DaRold:2005mxj, Falkowski:2008fz, Gherghetta:2009ac}, bulk/shear viscosities \cite{Kovtun:2003wp, Buchel:2003tz, Buchel:2007mf, Kapusta:2008ng}, and dynamical confinement \cite{Batell:2008zm}. 

Recently two of us proposed a dynamical two field model in order to reproduce the scalar glueball mass spectrum determined by lattice calculations at zero temperature \cite{Bartz:2018nzn,Meyer:2004gx}. After incorporating backreactions, we found a tachyonic mode in the spectrum. To regulate this mode, we introduced a Randall--Sundrum (RS) 3-brane with nonzero tension which acted as a mode cutoff in the putative dual field theory \cite{Randall:1999ee,Randall:1999vf,Rattazzi:2000hs,ArkaniHamed:2000ds}. The junction conditions at the brane regularized this mode and allowed for a good description of the scalar glueball mass spectrum.

At finite temperature, few extensions to the canonical AdS$_5$/CFT$_4$ program are analytically tractable. A certain subset, referred to as Einstein-dilaton models, permits a single bulk scalar field, making it amenable to analytically extract rich thermodynamic behavior \cite{Gursoy:2008za,Miranda:2009uw,Miranda:2009qp,Yaresko:2015ysa,Zollner:2018uep}. In these models, the finite temperature supergravity background is often fixed to be SAdS$_5$ or thermal AdS, and the spectrum of bulk matter is calculated in the absence of backreactions. The aim of this work is to provide an analytically tractable, solvable model which incorporates changes induced by a codimension one 3-brane and a bulk scalar field on the thermodynamics of bulk gravity at finite temperature.

We study Einstein-dilaton gravity with a bulk RS 3-brane situated at the $\mathbb{R}/\mathbb{Z}_2$ orbifold fixed point with finite tension. The tension of the codimension one brane combined with the dilaton provides a rich structure. In Sec. \ref{sec:action} we show that this model is sufficiently constrained to provide a unique generating superpotential. We solve the resulting nonlinear coupled equations of motion exactly in Sec. \ref{sec:solve}.  We identify the range of parameters which admit stable black brane formation in Sec. \ref{sec:para}.  We find the formation of black branes is contingent upon a simple lower bound 
\begin{equation}
\omega = \frac{3}{4 \pi} \frac{k}{\sigma G_5} > 1
\label{eqn:omega}
\end{equation}
where $\sigma>0$ refers to the brane tension, $k^2$ is proportional to the bulk cosmological constant $\Lambda$, $G_5$ is the bulk gravitational constant. When the bound is satisfied, black branes simultaneously form on either side of the RS 3-brane.  When the parameter $\lambda$ (to be defined later) takes the value 0, the scalar field decouples and the pure Randall--Sundrum model is recovered.  When $\lambda = \pm 1$ the scalar field may be identified as a tachyon, and black branes exist only above a certain minimum temperature.

In Sec. \ref{sec:thermo} we explore the thermodynamics of these finite temperature solutions. We find that, as long as the RS 3-brane is present, the black brane entropy is bounded above by
\begin{equation}
0 \le S < \frac{V_3}{4G_5}
\label{eqn:entropybound}
\end{equation} 
where $V_3 = \text{Vol}(\mathbb{R}^3)$.	The upper bound is saturated only in the limit of a vanishing brane tension, $\sigma\to 0$.  Examples of other systems which have a limiting entropy are recalled in Sec. \ref{sec:limit}.  As shown in Sec. \ref{sec:SUSY}, carefully taking the limit of vanishingly small tension and cosmological constant effectively reduces the model to Type IIB string theory on AdS$_5\times S^5$ of which the conjectured dual field theory is $\mathcal{N}=4$ SYM on $S^1\times\mathbb{R}^3$ at finite temperature with broken supersymmetry.  

Our conclusions are presented in Sec. \ref{sec:conclude}.  The results presented in this paper may have applications in holographic QCD, physics beyond the standard model, and models of the early universe.

\section{Action and Generating Superpotential}
\label{sec:action}

We will study the following $5d$ Einstein-Hilbert action minimally coupled to a single dimensionless bulk scalar field,
\be
\mathcal{S} = 2M^3\int d^5x \sqrt{|g|}\left[{R} + 12 k^2 - \frac{1}{2} (\partial\phi)^2 - V(\phi) \right] \, .
\label{eqn:action}
\ee
The (negative) cosmological constant is $\Lambda = - 6 k^2$, $2M^3=(16\pi G_5)^{-1}$ with $G_5$ being the bulk gravitational constant, and 
$(\partial\phi)^2 = g^{MN} (\partial_M \phi) (\partial_N \phi)$. Conformal invariance is broken by the scalar field.

In addition to the bulk action, we introduce an RS 3-brane $\Sigma$ with tension $\sigma$ in the following way. We extend the compactification radius of the ${S}_{r_c}^1/\mathbb{Z}_2$ orbifold to the real line \cite{Randall:1999ee}. The RS 3-brane is then situated at the orbifold fixed point of $\mathbb{R}/\mathbb{Z}_2$, and we expect to recover $4d$ Poincar\'e invariance at $\Sigma$. Let $\gamma = g_{\Sigma}$ be the induced metric; then we work with the boundary action
\be
\mathcal{S}_{\partial} = \mathcal{S}_{\text{GHY}} - \int_{\Sigma} d^4 x \sqrt{|\gamma|}\sigma.
\ee
where $\mathcal{S}_{\text{GHY}}$ is the Gibbons-Hawking-York boundary term which is required for a well-posed variational principle. Often the background needs to be identified and subtracted in order to render the total action finite. Here the presence of the RS brane provides a natural counterterm which, as noted in \cite{Emparan:1999pm}, we identify as the surface term responsible for yielding a finite action. 

\begin{figure}
\centering
\includegraphics[width=.5\textwidth]{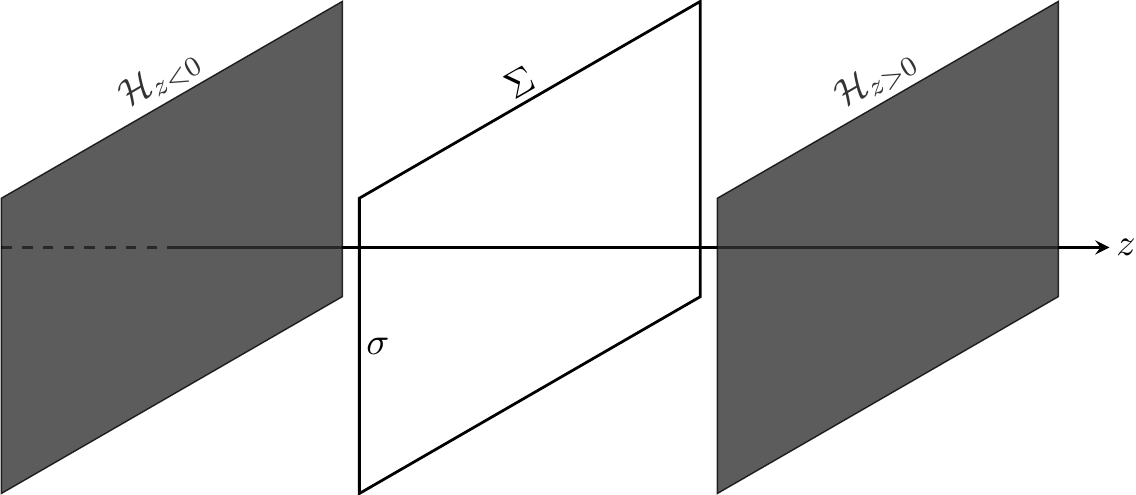}
\caption{The setup: two horizons form equidistantly on either side of $\Sigma$ when the tension $\sigma$ satisfies the bound $\omega>1$.}
\label{fig:D3BBsetup}
\end{figure}

To search for static black brane/thermal solutions, we employ the domain wall ansatz for the bulk metric
\begin{equation}
ds^2 = e^{-2A(y)}\left[-f(y){dt^2} + d\boldsymbol{x}^2 \right] + \frac{dy^2}{f(y)} \,.
\end{equation}
Here $f(y)$ is the blackening function whose zero indicates a horizon and $A(y)$ is the warp function. In these coordinates, $\Sigma$ is located at $y=0$, and junction conditions on $A(y)$ and $f(y)$ will have to be imposed.  See Fig. \ref{fig:D3BBsetup}. 
Note that in the special case where the RS brane is absent and $f(y)=1$, this metric arises from considering the near horizon asymptotics of $N_c$ coincident D3 branes and is equivalent to a warped AdS$_5$ with the deep IR located at $y\to\infty$ (see \cite{Gubser:1998bc,Karch:2006pv,Gherghetta:2009ac,Kapusta:2008ng,Bartz:2018nzn} for motivations and references therein).  

At zero temperature, the RS brane exhibits a $\mathbb{Z}_2$ orbifold symmetry which is expected to be preserved at finite temperature \cite{Bunk:2017fic}\footnote{In Ref. \cite{Park:2001nf} a perturbative analysis suggested an explicit breaking. However, the equations of motion are not linearly independent as the authors have claimed. Thus their analysis of the orbifold symmetry breaking is suspect.}. We search for solutions which respect the orbifold symmetry. A straightforward calculation of the Ricci tensor and the Ricci scalar yields
\ba
R_{tt} &=& f \left[\frac{1}{2} f'' - f A'' + 4 f A'^2 - 3f' A' \right] {e}^{-2A} \nonumber \\
R_{\boldsymbol{x}\boldsymbol{x}} &=& \left[f'A' - 4fA'^2 + fA'' \right] {e}^{-2A} \nonumber \\
R_{yy} &=& 4A{''} - 4A'^2 + \frac{3f'}{f} A' - \frac{f''}{2f} \nonumber \\
R &=& 8fA'' - 20 f A'^2 + 9 f' A' -  f''
\ea
where the prime notation denotes a derivative with respect to $y$. The resulting linearly independent equations of motion are
\ba
6 A'' &=& \phi'^2 + 12 q \delta(y) \label{eqn:A''} \\
12 f A'^2 - 3 f' A' &=& \frac{f}{2} \phi'^2 + 12 k^2 - {V} \label{eqn:A'} \\
(f\, \phi')'  - 4 f A' \phi' &=& \frac{dV}{d\phi} \,.
\ea
where $q \equiv {\sigma}/{24M^3}$. In the Randall-Sundrum model the vacuum solution requires $q = k$.  The first equation may be solved provided that there exists a continuous function $B(y)$ satisfying the following system of differential equations
\ba
A' &=& B + q[\theta(y) - \theta(-y)] \label{eqn:B} \\
B' &=& \frac{1}{6} \phi'^2 \,.
\ea
The Heaviside functions appear as a consequence of respecting the orbifold symmetry. Moreover, the delta function may be interpreted as a domain wall in the gravitational theory or, equivalently, as a mode cutoff in the putative dual field theory. 

In the rest of this work we look for solutions to the following coupled nonlinear differential equations,
\ba
B' &=& \frac{1}{6} \phi'^2 \label{eqn:B'} \label{eq:set1} \\
V - 12 k^2 &=& 3\left[f(B+q) \right]^\prime - 12 f(B+q)^2 \label{eqn:potential} \label{eq:set2} \\
\frac{dV}{d\phi} &=& (f \phi')' - 4 f (B+q) \phi' \label{eq:set3}
\ea
where we have eliminated the warp function in favor of $B(y)$.  Although we have restricted to $y>0$, the deformation $y \to |y|$ and 
$q \to q\,\Theta(y)$, where $\Theta(y) \equiv \theta(y)-\theta(-y)$, recovers the orbifold symmetry.  We still need to find the junction conditions for these functions on $\Sigma$.

A common issue with Einstein-dilaton theories is the internal consistency of the non-linear differential equations. To remedy this issue, we construct the following algorithm which solves the system uniquely up to boundary conditions on the 3-brane.
\begin{enumerate}
\item Introduce $W(\phi)$ as a superpotential \cite{DeWolfe:1999cp,Skenderis:1999mm} which generates the potential $V(\phi)$.
\item Promote the metric function $B(y)$ to $B(\phi)$.
\item Introduce a function $D(\phi)\equiv e^{-a\phi} \left[ B(\phi)+q \right]$.
\item Transform the coordinates via  $du = e^{a\phi(y)} dy$.
\item Impose consistency on Eqs. (\ref{eq:set2}) and (\ref{eq:set3}).
\end{enumerate}
	
Superpotential methods are often used to introduce a holographic RG flow from the UV to the IR along the bulk direction and reduce the order of the differential equations \cite{Gursoy:2008za,Kiritsis:2016kog}. These auxiliary superpotentials are tasked to generate the potential at zero temperature by a first order gradient flow along $\phi$ via 
\be
V(\phi) = 18\left(\frac{dW}{d\phi}\right)^2 - 12 W^2 - 24 k W\Theta(y) \,,
\label{eqn:superpotential}
\ee
where the discontinuity $\Theta(y)$ arises from the junction condition \cite{Bartz:2018nzn}. At zero temperature, the superpotential is identified with the metric function $B(y)$. At finite temperature, the situation is more subtle. Though the relation \eqref{eqn:superpotential} holds at finite temperature since the potential should be temperature independent, the coupling between $f$ and $B$ via Eqs. \eqref{eq:set1} to \eqref{eq:set3} breaks the identification between $W(\phi)$ and $B(\phi)$. In this case, $W(\phi)\propto B(\phi)$ with the factor being set by the brane tension $\sigma$ and the bulk constants $\Lambda$ and $G_5$. 

While the superpotential $W(\phi)$ uniquely generates the potential $V(\phi)$, the function $B(\phi)$ reduces the non-linear equation \eqref{eqn:B'} to a first order gradient flow along $\phi$
\be
\frac{dB}{d\phi} = \frac{1}{6}\frac{d\phi}{dy} \,,
\ee
provided that $\phi'$ has no critical points in the bulk.  A unique solution for the dilaton may be constructed from 
\be
\frac{dD}{d\phi} + a D = \frac{1}{6} \frac{d\phi}{du} \,.
\label{eqn:phi-u}
\ee
We henceforth set the parameter $a=1/\sqrt{6}$. This value is determined by the conformal transformation from the string to Einstein frame in order to obtain a canonical kinetic term \cite{Batell:2008zm} and separately was found to give rise to linear Regge trajectories in certain AdS/QCD contexts \cite{Meyer:2004gx,Erlich:2005qh}.

Next, note that both Eqs. \eqref{eq:set2} and \eqref{eq:set3} are differential equations containing $f$ and $f'$.  Steps 3 and 4 of the algorithm reduce to
\be
18 D \left[ \frac{dD}{d\phi} + a D \right] \frac{df}{d\phi} + 9 \left[ 2 \left(\frac{dD}{d\phi}\right)^2 + 4a D \frac{dD}{d\phi} - D^2 \right] f =
\left( V - 12 k^2 \right) e^{-2a\phi}
\label{Ephi12}
\ee
and
\be
36 \left[ \frac{dD}{d\phi} + a D \right]^2 \frac{df}{d\phi} + 18 \left[ \frac{dD}{d\phi} + a D \right] 
\left[ 2 \frac{d^2D}{d\phi^2} + 4a \frac{dD}{d\phi} - D\right] f = \frac{dV}{d\phi} \, e^{-2a\phi} \,.
\label{Ephi32}
\ee
One may ask whether these two equations are consistent with each other.  These can be put in the form
\ba
f + P_1  \frac{df}{d\phi} &=& Q_1 \nonumber \\
f + P_2  \frac{df}{d\phi} &=& Q_2 \,.
\ea
Then $P_1 = P_1$ results in the differential equation
\be
D \frac{d^2D}{d\phi^2} = \left( \frac{dD}{d\phi} \right)^2
\ee
whose only solution is
\be
D = D_0 \exp((\lambda - 1)a \phi)  
\ee
where $D_0$ and $\lambda$ are any real constants.  Setting $Q_1 = Q_2$ results in
\be
\frac{d}{d\phi} \left( V - 12k^2 \right) = 2 \lambda a \left( V - 12k^2 \right)
\ee
with solution
\be 
V - 12k^2 = V_0 \, e^{2 \lambda a \phi}
\label{onlyV}
\ee
We may conclude that expression (\ref{onlyV}) is the only potential for a single scalar field that admits a black brane.  Note however that a constant blackening function $f = 1$ places no restrictions on the potential because, after multiplication by $e^{2a\phi}$, Eq. (\ref{Ephi32}) just becomes the derivative with respect to $\phi$ of Eq. (\ref{Ephi12}).

The superpotential which generates this $V$ and satisfies the junction condition is
\be
W = k \left( e^{\lambda a \phi} - 1 \right) \Theta(y)
\ee
where we used the fact that $\Theta^2(y)$ is continuous at $y=0$.  It follows that $V_0 = 3 ( \lambda^2 - 4) k^2$.  It also follows that
\be
B = q \left( e^{\lambda a \phi} - 1 \right) \Theta(y)
\ee
so that $W = \omega B$ where $\omega \equiv k/q$.

\section{Exact Solutions}
\label{sec:solve}

Using the solutions $D(\phi)$ and $V(\phi)$ found in the previous section, the functions $P(\phi)$ and $Q(\phi)$ reduce to constants
\ba
P &=& - \frac{6a\lambda}{4 - \lambda^2} \nonumber \\
Q &=& \frac{k^2}{q^2} = \omega^2 \,.
\ea
Then the differential equation for the blackening function has a simple closed form
\be
f(\phi) = \omega^2 - \left(\omega^2 -1\right) e^{-\phi/P} \,.
\ee
The dilaton profile may be obtained from Eq. \eqref{eqn:phi-u}. The solution in the original $y$ coordinate is
\begin{equation}
\phi(y) = - \frac{1}{a\lambda}\ln\left(1-q\lambda^2 |y| \right) \,.
\label{eqn:phi-profile}
\end{equation}
Equipped with the profile $\phi(y)$, the functions $f$, $V$ $B$, and by extension $A$, obtain analytic forms in terms of $y$. Thus we have uniquely solved the system of Eqs. \eqref{eq:set1} to \eqref{eq:set3} incorporating the backreactions of the scalar field and the codimension one hypersurface on the background geometry.

In order to facilitate the analysis of the parameter space of solutions, we perform a coordinate transformation $dy = e^{-A(y(z))} dz$ where
\begin{equation}
A(y) = - \frac{1}{\lambda^2} \ln \left(1-q\lambda^2 |y|\right)
\label{eqn:Ay-profile}\\
\end{equation}
is the solution to Eq. \eqref{eqn:B}. The metric then takes the canonical thermal form with an overall conformal (warp) factor
\begin{equation}
ds^2 = e^{-2 A(y(z))} \left[-f(z) dt^2 + d\boldsymbol{x}^2 + \frac{dz^2}{f(z)} \right]
\label{eqn:metricz}
\end{equation}
with the coordinates related by
\begin{equation}
1 - \lambda^2 q |y(z)| = \left[1+ (1-\lambda^2) q |z| \right]^{-\lambda^2/(1-\lambda^2)} \,.
\end{equation}
In these coordinates the functions respecting the orbifold symmetry are 
\ba
f(z) &=& \omega^2 - (\omega^2 -1) \left[1 + (1-\lambda^2) q |z| \right]^{(4-\lambda^2)/(1-\lambda^2)} \label{eqn:fz-profile} \\
A(z) &=& \frac{1}{1-\lambda^2} \ln \left[1 + (1-\lambda^2) q |z| \right]  \label{eqn:Az-profile} \\
\phi(z) &=& \frac{\lambda}{a (1-\lambda^2)} \ln \left[1 + (1-\lambda^2) q |z| \right] \label{eqn:phiz-profile} \\
V(z) &=& 12 k^2 - 3(4-\lambda^2) k^2 \left[1 + (1-\lambda^2) q |z| \right]^{2\lambda^2/(1-\lambda^2)} \label{eqn:Vz-profile} \,.
\ea

\section{Parameter Space Constraints}
\label{sec:para}

We now consider the conditions in which a horizon forms and study the parameter space of finite temperature solutions. Heuristically, $\lambda$ characterizes the dilaton content while $\omega$ parameterizes the metric content of the theory. We will study the parameter space generated by $(\lambda,\omega)$ and illustrate certain extremal values in preparation for a finite temperature analysis. 

To constrain $\lambda$, note that the dilaton profile contains a logarithmic singularity at finite $z$ when $|\lambda| > 1$. To prevent this, the relevant domain which provides nonsingular $\phi$ and real $V(\phi)$ for all $z$ is given by $-1 \le \lambda \le 1$. Though the edge cases of the parameter domain of $\lambda$ have to be studied with care, the limits of the functions $f, A, \phi$ and $V$ exist and respect the orbifold symmetry. 

First, consider the limit $\lambda \to 0$ when the dilaton decouples
\ba
\phi(z) &=& 0 \nonumber \\
V(z) &=& 0 \nonumber \\
A(z) &=& \log(1+q|z|)  \nonumber \\
f(z) &=& \omega^2 - (\omega^2 -1)(1+q|z|)^4  \label{eqn:vacuumlimit}  \,,
\ea
so that only the metric degrees of freedom remain. In this case, we see that the action reduces to an Einstein-Hilbert action with a cosmological constant and a boundary term from the brane tension. This limit will provide useful thermodynamic consistency checks between this spacetime and AdS$_5$.

Next, we have the limit where $\lambda \to \pm1$
\ba
\phi(z) &=& \pm \sqrt{6}q|z| \nonumber \\
V(z) &=& 3 k^2 (4 -3e^{2q|z|}) \nonumber \\
A(z) &=& q|z| \nonumber \\
f(z) &=& \omega^2 - (\omega^2 -1)e^{3q |z|}  \,.
\label{eqn:tachyonlimit}
\ea
Remarkably, we find that the dilaton reproduces the tachyonic profile considered in Ref. \cite{Batell:2008zm}. This should be expected as single field AdS/QCD models with a hard wall have a spectrum of squared masses which grow as $m_n^2\propto n^2$ for high excitation number (see Ref. \cite{Karch:2006pv} for a novel discussion). Effectively, the 3-brane confines the excitations of the dilaton, analogous to a Schr\"odinger equation for a particle in a box. To reproduce the linear Regge trajectories found in AdS/QCD models, the relationship $A = (a/\lambda)\phi$ must be broken, which can be done through additional bulk fields or boundaries, modified junction conditions, or in the vacuum limit.

\begin{figure}[th]
\center{\includegraphics[width=290pt]{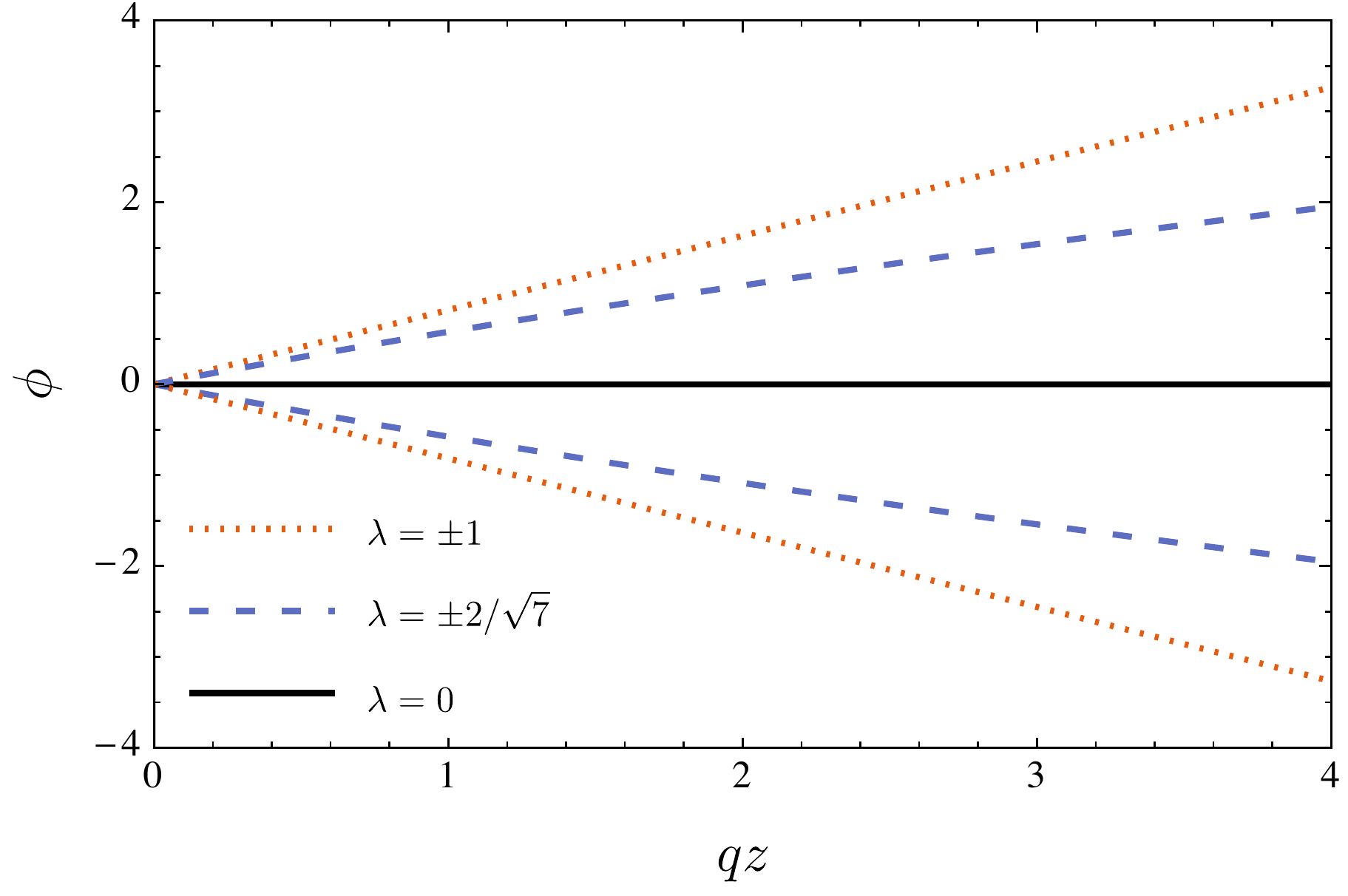}}
\caption{The scalar field as a function of the dimensionless variable $qz$ for representative values of $-1 \le \lambda \le 1$.}
\label{DilatonProfiles}
\end{figure}

\begin{figure}[th]
\center{\includegraphics[width=290pt]{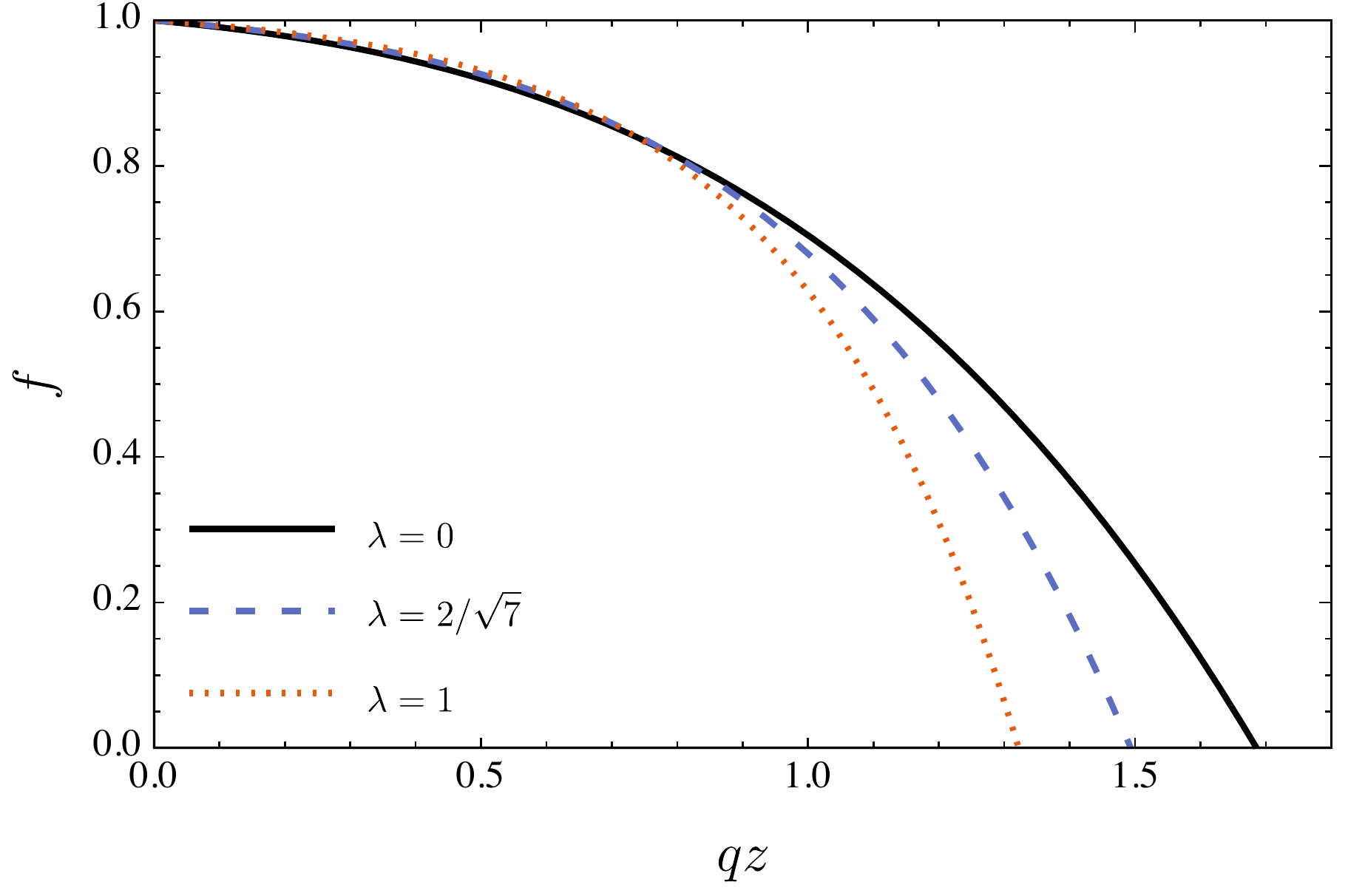}}
\caption{The blackening function $f$ as a function of the dimensionless variable $qz$ for representative values of $\lambda$.}
\label{fProfiles}
\end{figure}

For a tension $\sigma>0$, the first constraint we may impose is $\omega > 0$. The coefficient $1-\omega^2$ in Eq. \eqref{eqn:fz-profile} then determines the existence of a black brane. For $0 < \omega < 1$ ($\omega>1)$, $f(z)$ is a monotonically increasing (decreasing) function with respect to $|z|$. Hence, for a real solution to $f(z_h)=0$ at some $|z_h|>0$ to exist, the tension must satisfy $\omega > 1$.  As a consequence of the orbifold symmetry, when this condition is satisfied, two black branes surrounding the 3-brane form.  Alternatively, this bound may be interpreted as the necessary condition\footnote{We thank Robert Myers for bringing this to our attention.} for the system to be in equilibrium. Furthermore, one can check that the Kretschmann invariant $R_{abcd}R^{abcd}$ is finite at $|z_h|$.  See Figs. \ref{DilatonProfiles} and \ref{fProfiles}.

The second set of solutions are extremal, i.e., when the tension of the 3-brane is such that $q = k$. As $q \to k$ from below, the horizons $\mathcal{H}_{z < 0}$ and $\mathcal{H}_{z > 0}$ are pushed to their respective asymptotics, and black brane formation is prohibited when the equality holds. As we will subsequently show, in the extremal case $T=0$, and these solutions are similar to AdS in Poincar\'e coordinates with the Killing horizon replaced with the 3-brane.  

Lastly, we remark that solutions with $\omega<1$ are forbidden as valid thermodynamic solutions. In these cases, $f$ diverges as $|z|\to\infty$. Without a horizon to contain these singularities, these spacetimes are wormhole--like solutions and thus are excluded from our finite temperature study.

\section{Thermodynamics}
\label{sec:thermo}

In this section the entropy is calculated from the Bekenstein-Hawking formula with the temperature identified from the periodicity of Euclidean time \cite{Gibbons:1976pt,Bekenstein:1973ur,Hawking:1976de}. 

We begin by deriving the Hawking temperature associated with the metric in $z$ coordinates Eq. \eqref{eqn:metricz}. We initially focus on theories which are non-tachyonic, i.e., $|\lambda|<1$. Since $\omega>1$ in order for black branes to exist, it is convenient to solve $f(z_h)=0$, Eq. \eqref{eqn:fz-profile}, for $z_h$ and relate it to the tension via
\be
\left[1 + (1-\lambda^2) q |z_h| \right]^{(4-\lambda^2)/(1-\lambda^2)} = \frac{\omega^2}{\omega^2-1} = \left[1-q^2/k^2 \right]^{-1}
\label{eqn:horizon}
\ee
where we have used the definition of $\omega$. Note that as $q\to k$ from below, the location of the horizons are taken to their respective asymptotics, $|z_h| \to \infty$. This justifies the explanation given earlier. Further, note that black brane formation is prohibited in the extremal case $\omega=1$ since $f(z)=1$. 

From the periodicity in Euclidean time, the Hawking temperature reads
\be
T = - \frac{1}{4\pi} \dot{f}(z_h) = \frac{k}{4\pi} (4 - \lambda^2) \left( \frac{\omega^2 - 1}{\omega} \right)
\left( \frac{\omega^2}{\omega^2 - 1} \right)^{3/(4-\lambda^2)}
\ee
\begin{figure}[th]
\center{\includegraphics[width=290pt]{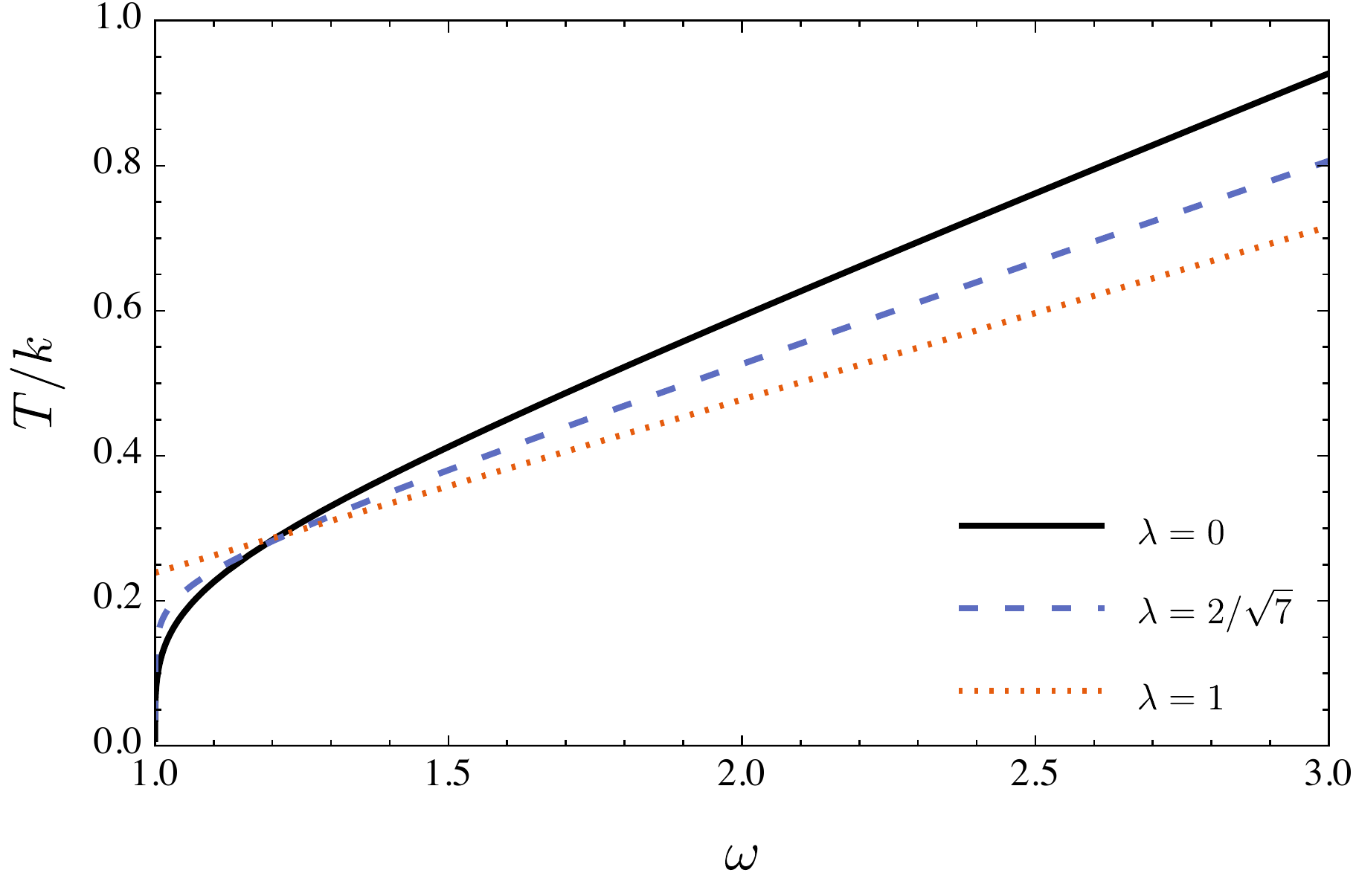}}
\caption{The temperature in units of $k$ as a function of the dimensionless variable $\omega$ for several values of $\lambda$.  It depends only on $|\lambda|$ and not on its sign.  For $|\lambda| = 1$ there is a minimum temperature below which a black brane solution emerges.}
\label{TempProfiles}
\end{figure}
Note that the extremal case $q=k$ precisely corresponds to $T=0$ unless the theory has tachyonic scalars, in which a minimum temperature develops at $T_{\min}=3k/4\pi$.  See Fig. \ref{TempProfiles}.

\begin{figure}[th]
\center{\includegraphics[width=290pt]{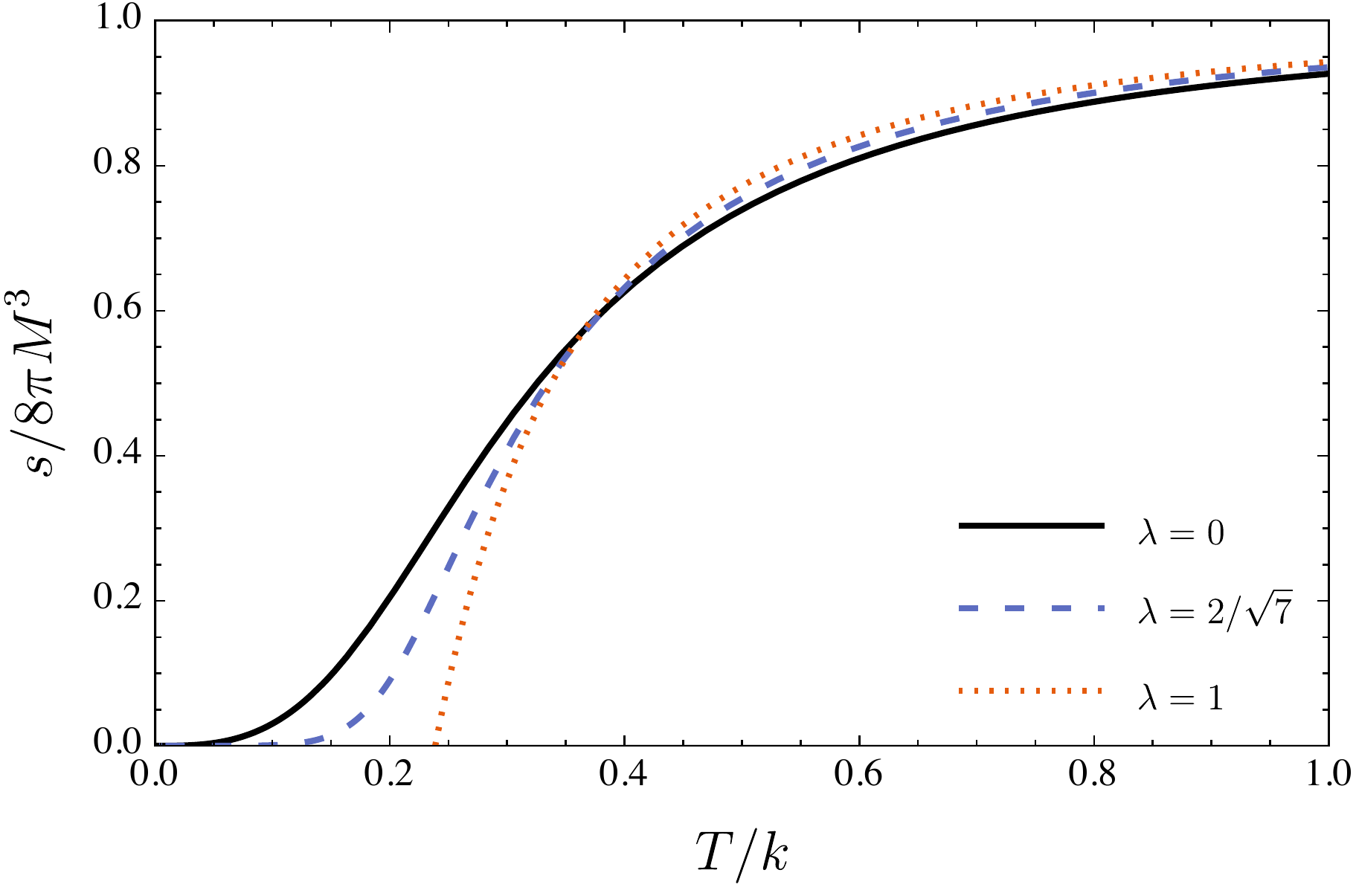}}
\caption{The entropy density, in units of $8 \pi M^3$, as a function of the temperature, in units of $k$, for several values of $\lambda$.  It depends only on 
$|\lambda|$ and not on its sign.  For $|\lambda| = 1$ the entropy goes to zero at a non-zero temperature.}
\label{entropyprofile}
\end{figure}

The entropy can be found from the Bekenstein-Hawking area law $S = \text{Area}(\mathcal{H})/4G_5$ \cite{Hawking:1976de,Bekenstein:1973ur}.  For the area of the horizon we have $dt = dz = 0$.  The area is computed from $\text{Area}(\mathcal{H}) = \int \sqrt{g_3(z_h)} \, d^3x$, where $g_3$ is the determinant of the metric for the 3-dimensional space.  It is
\be
 g_3(z_h) = {\rm e}^{-6A(z_h)} = \left( 1 - \frac{q^2}{k^2} \right)^{6/(4-\lambda^2)}
\ee
The entropy density is therefore
\be
s = 8 \pi M^3 \left( \frac{\omega^2 - 1}{\omega^2} \right)^{3/(4-\lambda^2)}
\ee
Interestingly, the entropy density is bounded from above: $0 \le T < \infty$ and $0 \le s < 8 \pi M^3$.  See Fig. \ref{entropyprofile}.
In the low temperature limit the entropy grows like $s \sim M^3 (T/k)^{3/(1-\lambda^2)}$.  

\section{Examples of Systems with Limited Entropy}
\label{sec:limit}

One may wonder how unusual it is for a system to have a limiting entropy with increasing temperature.  It typically happens when two conditions are met: There is a restriction on the allowed phase space, and there is a relevant energy scale apart from the temperature.  We will illustrate this with two examples.  The first involves a restriction on the allowed energies, and the second involves a restriction on the volume available to a gas of extended particles.

\subsection{Two level system of fermions}

Consider a system of $N$ noninteracting fermions, each of which can be in a state with zero energy or energy $E_0$.  The partition function is
\be
\ln Z = N \ln \left( 1 + e^{-E_0/T} \right) \,.
\ee
The total entropy is
\be
S =  \frac{d}{dT} \left( T \ln Z \right) = \ln Z + N \, \frac{E_0/T}{ e^{E_0/T} + 1} \,.
\ee
When $T/E_0 \to \infty$, $S \to N \ln 2$.  This is well known.  Since the entropy is the logarithm of the number of available states, and at high temperature the two states are equally occupied, one obtains the $\ln 2$.

Next, consider a gas of massless, noninteracting spin $\thalf$ fermions (including anti-particles) which have a maximum allowed momentum $p_0$.  This is analogous to a gas of phonons in a solid.  The entropy density is
\be
s = \frac{2T^3}{\pi^2} \int_0^{p_0/T} dx \, x^2 \left[ - n \ln n - (1-n) \ln (1-n) \right]
\ee
where the occupation number is $n = 1/(e^x+1)$.  When $T \ll p_0$ the entropy density goes to $s \to (7 \pi^2/45) T^3$.  When $T \gg p_0$ it goes to 
$s \to p_0^3 \, (2\ln 2)/(3\pi^2)$.

When bosons are considered instead of fermions, the entropy grows logarithmically as $\ln (E_0/T)$ and $\ln (p_0/T)$, respectively.  The difference, of course, is the fact that an arbitrary number of bosons can occupy any given quantum state.  

\subsection{Excluded volume model of a hadron gas}

Another example is a Van der Waals type of excluded volume model for a hadron gas.  In one version of the model, the volume occupied by a hadron is approximated by $E(p)/\epsilon_0$, where $E(p)$ is the energy of the hadron with momentum $p$ and $\epsilon_0$ is a constant with units of energy/volume \cite{Kapusta:1982qd, Kapusta:1989tk}.  The equation of state is expressed in parametric form.
\ba
P(T) &=& \frac{P_{\rm pt}(T_*)}{1 - P_{\rm pt}(T_*)/\epsilon_0} \nonumber \\
s(T) &=& \frac{s_{\rm pt}(T_*)}{1 + \epsilon_{\rm pt}(T_*)/\epsilon_0} \nonumber \\
\epsilon(T) &=& \frac{\epsilon_{\rm pt}(T_*)}{1 + \epsilon_{\rm pt}(T_*)/\epsilon_0}
\ea
The subscript ``pt" refers to the pressure, entropy, or energy density of a gas of hadrons treated as point particles.  The temperature is obtained from the parameter $T_*$ by the formula
\be
T = \frac{T_*}{1 - P_{\rm pt}(T_*)/\epsilon_0} \,.
\ee
It can be readily verified that the thermodynamic identities $s = dP/dT$ and $\epsilon = -P + Ts$ are satisifed.  

The temperature $T \to \infty$ at a finite value of $T_*$ determined by $P_{\rm pt}(T_{* {\rm max}}) = \epsilon_0$.  The pressure $P$ is unbounded but the entropy and energy densities have finite upper limits.  Consider, for example, a gas of massless bosons and fermions with the equation of state $P_{\rm pt}(T_*) = c T_*^4$.  Then, when the excluded volumes are taken into account, the entropy density is
\be
s(T) = \frac{4 c T_*^3}{1 + 3 c T_*^4/\epsilon_0}
\ee
with
\be
T = \frac{T_*}{1 - c T_*^4/\epsilon_0} \,.
\ee
Thus $T_{* {\rm max}} = (\epsilon_0/c)^{1/4}$ and
\be
s(T) < c (\epsilon_0/c)^{3/4} \,.
\ee
On the other hand, when $T \ll T_{* {\rm max}}$, the entropy density is unaffected by the excluded volume, and $s \to 4 c T^3$.  The reason for an upper limit on the entropy in this excluded volume model is the restriction in coordinate space, not momentum space.

\section{Recovering AdS/CFT}
\label{sec:SUSY}

In order to make comparison with SAdS$_5$ black branes, we write the metric in the form
\be
ds^2 = H^{-1/2}(r) \left[-f(r){dt^2} + d\boldsymbol{x}^2 \right] + H^{1/2}(r) \frac{dr^2}{f(r)} \,.
\ee
There is no scalar field so $\lambda = 0$.  The $z$ and $r$ coordinates are related by
\be
H = 1 + \frac{L^4}{r^4} = (1 + qz)^4 \,.
\ee
In the far IR ($z \to \infty$, $r \to 0$) we identify $L = 1/q$.
The blackening and warp functions become
\ba
f(r) &=& 1-\left(\frac{r_h}{r} \right)^4 \\
A(r) &=& \frac{1}{4} \ln\left( 1 + \frac{L^4}{r^4} \right)
\ea
where the warp function may be identified as an overall conformal factor. The horizon is given by 
\be
\frac{r_h}{L} = \left( \frac{\omega^2 - 1}{\omega^2} \right)^{1/4} \,.
\ee
In the near horizon limit the metric takes the form
\begin{equation}
d s^2 = \left(\frac{r}{L}\right)^2 \left[-f(r) dt^2 + d\boldsymbol{x}^2 + 
\left(\frac{L}{r} \right)^4 \frac{d r^2}{f(r)} \right].
\end{equation}
The finite temperature AdS/CFT result is obtained by taking the limit $\omega \to 1$ with $r_h$ held fixed.  In a sense this is a low temperature limit because, with 
$L = 1/q \to \infty$, we have $r_h/L \to \pi T L \to (\omega^2 -1)^{1/4}$.  The AdS/CFT dictionary says that
\be
8 \pi M^3 = \frac{1}{4 G_5} \equiv \frac{N_c^2}{2\pi L^3}
\ee
where $N_c$ is the number of colors of the gauge field.  Hence the entropy density becomes
\be
s = \frac{\pi^2 N_c^2}{2} T^3
\ee
which is the canonical result.

\section{Conclusion}
\label{sec:conclude}

In this paper we studied Einstein-dilaton gravity with a bulk Randall--Sundrum 3-brane situated at the $\mathbb{R}/\mathbb{Z}_2$ orbifold fixed point.  We looked for solutions to the equations of motion which incorporated a black brane, including backreaction of the dilaton field on the metric.  A black brane can only form if the potential has a particular functional form.  With this potential we discovered exact solutions to the equations of motion.  The temperature and entropy density are readily determined from these solutions.  The temperature is controlled by varying the tension on the RS 3-brane.  It turns out that the entropy has an upper bound given simply by $\text{Vol}(\mathbb{R}^3)/4G_5$.  Examples of other systems which also have a limiting entropy were provided.  Typically they require some restriction in the allowed phase space along with a relevant dimensional parameter, apart from the temperature, to provide a scale.

Further lines of inquiry naturally present themselves.  Why is it that the functional form of the potential is restricted in order that a black brane can be formed?  Formation of a black brane and its associated Hawking temperature and Bekenstein entropy, following from the area law, are the usual means to study such theories.  There is no analogous restriction at zero temperature.  

Extensions of the model to include more scalar fields, such as glueball and chiral fields, might provide better descriptions of pure Yang--Mills theory and QCD.  Extensions to include physics beyond the standard model of particle physics and general relativity might be very relevant to the physics of the early universe.  

\section*{Acknowledgments}
We thank T. Gherghetta for comments on the manuscript. This work was supported by the U.S. DOE Grant No. DE-FG02-87ER40328.  A. D. is supported
by the National Science Foundation Graduate Research Fellowship Program under Grant No. 00039202.  C. P. acknowledges support from the CLASH project (KAW 2017-0036).

\bibliography{KDPPaper_v5}

\end{document}